\documentclass{article}
\usepackage{ijcai22}

\usepackage{times}
\usepackage{soul}
\usepackage{url}
\usepackage{cite}
\usepackage[hidelinks]{hyperref}
\usepackage[utf8]{inputenc}
\usepackage[small]{caption}
\usepackage{graphicx}
\usepackage{amsmath}
\usepackage{amsthm}
\usepackage{booktabs}
\usepackage{algorithm}
\usepackage{algorithmic}
\usepackage{natbib}
\title{Neighbor2vec: an efficient and effective method for Graph Embedding}

\author{
    Zhiming Lin
    \affiliations
    Tongdun
    \emails
    nklinzhiming@gmail.com
}

\begin{document}

\maketitle

\begin{abstract}
   Graph embedding techniques have  led to significant progress in recent years. However, present techniques are not effective enough to capture the patterns of networks. This paper propose neighbor2vec, a neighbor-based sampling strategy used algorithm to learn the neighborhood representations of node, a framework to gather the structure information by feature propagation between the node and its neighbors. We claim that neighbor2vec is a simple and effective approach to enhancing the scalability as well as equality of  graph embedding, and it breaks the limits of the existing  state-of-the-art unsupervised techniques. We conduct experiments on several node classification and link prediction tasks for networks such as ogbn-arxiv, ogbn-products, ogbn-proteins, ogbl-ppa,ogbl-collab and ogbl-citation2. The result shows that Neighbor2vec’s representations provide an average accuracy scores up to 6.8 percent higher than competing methods in node classification tasks and 3.0 percent higher in link prediction tasks. The neighbor2vec’s representations are able to outperform all baseline methods and two classical GNN models in all six experiments.
\end{abstract}

\section{Introduction}
There are many kinds of networks in real word including  social network, collaborative network, citation network, the World Wide Web etc. Different structures and properties make networks more complicated than the other forms of data, such as the language, the pictures etc. For example we got homogeneous network, heterogeneous network, static network and dynamic network. There are many important tasks and applications in networks ,such as community detection, link prediction, node classification, recommendation and  graph visualization. Taking a typical node classification for example,  The bank may be very interested in finding out those financial fraudsters, or in social networks we might be interested in predicting interests of users to offer better matched products or services. 

In supervised learning, any algorithm requires a set of features. A typical way to deal with this demand is engineering domain-specific features based on expert knowledge and statistical methods. However, with the growing tasks appearing, this way may encounter lots of  challenges.To handle the dilemma, graph embedding stands out and has been provided many promising and exciting results for various tasks and applications listed above. Numbers of recent research papers have proposed a similar network representation method based on word2vec~\cite{mikolov2013efficient} method, which first proposed in natural language processing (NLP) field. Such as DeepWalk~\cite{perozzi2014deepwalk}, LINE~\cite{tang2015line}, Item2Vec~\cite{barkan2016item2vec}, node2vec~\cite{grover2016node2vec}, metapath2vec~\cite{dong2017metapath2vec} etc. Instead of designing a specific model for a single task, these network-representation methods are designed to engineer useful and general features for various downstream tasks based on the network’s structure only. First they generate model inputs by a random or biased or specific walk. Then they design a model with the object seeking to preserve local neighbors of nodes, which is independent from the downstream tasks.  Finally, they train the model to get the embedding.

The walk-based sampling strategies, either random walk or guided walk, from where we stand, however,  do not represent the neighborhood relationship in a sufficient way. First, in NLP field, the words are treated to be connected with those who present within a fixed wind size of a natural language sentence~\cite{mikolov2013efficient}. But things get changed in a network that one node is easy to be influenced by its neighbors. The further the distance, the minor the influence. Second, we can see that the walk based sampling strategies sample approximately only two neighbors for one central node in one walk. And we think it is easily to be insufficient samples for those nodes who maintain large number of neighbors. Third, the sampling methods ignore the total structure of the network.  So is there any technical method to reuse the structure information to help to generate the  graph embedding?

In this paper we propose the neighbor2vec, a no-walk sampling strategy used algorithm, a simple and effective approach to enhancing the scalability as well as equality of unsupervised graph embedding method, a framework to gather the structure information by feature propagation between the node and the neighbors. First, intuitively, we gather the central node and its neighbors as the model inputs. Second, we optimize a neighbor-based objective function that maximize the likelihood of of preserving network neighborhoods of nodes motivated by prior work. Third, we aggregate the embedding vectors of its neighbors. Finally we update the center node’s embedding by a weighted sum of its original’s and the aggregating vectors. 

Overall, our paper makes the following contributions:

\begin{enumerate}
    \item We propose neighbor2vec a simple, high-quality and  efficient scalable graph embedding algorithm for neighborhood relationship representation in networks. Neighbor2vec focus on nearest neighbors and optimizes a neighbor-based objective function that maximize the likelihood of preserving network neighborhoods of nodes. The algorithm breaks the link rule of the walk-based algorithms and improves the effectiveness and efficiency of the representation. 
    \item The neighbor2vec allows for significantly more parallelization. We propose a novel network embedding model using stochastic gradient decent to optimize neighborhoods relationship representation. And we introduce a bi-directional influenced mechanism by propagating embedding to neighbors  as well as updating the embedding with the aggregating  information from the neighbors.
    \item We conduct extensive experiments on several real-world large-scale networks. We see the inspiring results on several real-word networks, and prove the effectiveness and efficiency of neighbor2vec.
\end{enumerate}

The rest of the paper is structured as follows: In Section 2, we briefly survey related work in unsupervised feature learning for networks. In Section 3, we present the technical details for unsupervised feature learning method of  neighbor2vec. In Section 4, we empirically evaluate neighbor2vec on prediction tasks over nodes and edges over  various real-world networks and assess the parameter sensitivity, perturbation analysis, and scalability aspects of our algorithm. In Section 5, we conclude with a discussion of the neighbor2vec framework and highlight some promising directions for future work.

\section{Related Work}

Graph embedding is a typical method of dimension reduction of network.  Several methods such as PCA~\cite{belkin2001laplacian}, MDS~\cite{roweis2000nonlinear}, IsoMap~\cite{tenenbaum2000global}, LLE~\cite{yan2006graph} have been proposed in last few years . These methods meet great challenges when scaling to large-scale networks because of  the complexity of which is at least quadratic to the number of nodes. These methods have lots of drawbacks on computational performance and statistical performance.

Another recent method is called graph factorization~\cite{ahmed2013distributed}. By using stochastic gradient descent based matrix factorization, it finds a way to get the embedding of a network. The object of graph factorization, however, is not suitable for network. Therefore graph factorization do not effectively preserve the structure of network. Besides graph factorization applies to the undirected networks only and suffers from directed network.

The most related works, to our best knowledge, are DeepWalk and node2vec, which are inspired by recent advancements in natural language process trying to get word representation in vector space. To be more detailed, they use skip-gram model to learn continuous representations for nodes by optimizing a neighborhood preserving likelihood objective using stochastic gradient descent by negative sampling~\cite{mikolov2013distributed}. DeepWalk and node2vec represent a network as a “document”  and then they generate “sentences” which are ordered sequences of nodes by different sampling strategies. Although they show  very good performance, we still here ask. 

Is the continuous walks really necessary for network since the skip-gram model is not position-aware at all?  Is there any more reasonable sampling strategy for networks except deep-first search strategy? Is there a way that we can pay more attention to explore the relationship between the nodes and neighbors? 

We try to answer this question by designing a novel  and efficient sampling strategy added with a simple bi-directional influenced mechanism between within neighbors. Our sampling strategy is breadth-first search and focus on node’s neighbors only, which is trivially parallelizable and more efficient.  We use skip-gram model to get the node embedding and then aggregate from the neighbors as well as propagate to the neighbors. The reason we design this method is that we intuitively  think nodes and neighbors can influence each other. Node can be influenced by its neighbors and vice versa. Several experiments have prove that this intuition  makes sense.

\section{neighbor2vec:The Neighbor Matters for Scalable Network Embedding}
Our analysis is general and applies to any (un)directed, (un)weighted network. We assume that we have an undirected network/graph $G = (V,E)$, where there are $\begin{smallmatrix}
\resizebox{.12\linewidth}{!}
{$n = \left| V \right|$}
\end{smallmatrix}$ nodes with features represented by a matrix $X \in R^{(n*p)}$ where p is the dimension of the matrix. The objective is to maximize the network probability in terms of local structures, that is:

\begin{equation}
\resizebox{.59\linewidth}{!}{
    $\displaystyle$
    $\mathop{\arg\max}\limits_{\theta}$
    $\prod \limits_{v \in V}$ 
    $\prod \limits_{c \in N(v)}$ 
    $p(c|v; \theta)$
}
\end{equation}%
where N(v) is the neighborhood of node v in the network G, which can be defined in different ways such as v’s one-hop neighbors, two-hop neighbors or both. And $\begin{smallmatrix} \resizebox{.15\linewidth}{!} {$p(c|v;\theta)$} \end{smallmatrix}$ defines the conditional probability of having a neighbor node c given a node v which is controlled by $\theta$. In order to make this optimization tractable , we make two assumptions:
\begin{enumerate}
    \item The likelihood of observing one of the neighbors is independent of observing any other neighbor.
    \item Symmetry in feature space. In feature space, both the node v and its neighbor   have a symmetry effect over each other. 
\end{enumerate}
Therefore, $\begin{smallmatrix} \resizebox{.15\linewidth}{!} {$p(c|v;\theta)$} \end{smallmatrix}$ is commonly defined as:

\begin{equation}
\resizebox{.59\linewidth}{!}{
    $\displaystyle
    p(c|v;\theta)  = \frac{exp(x_c*x_v)}{\sum \limits_{u \in N(v)}(exp(x_u*x_v)}$
}
\end{equation}

The algorithm consists of three main components:
\begin{enumerate}
    \item First,  a neighbor generator  takes a graph G and samples uniformly a fixed number of a node’s neighbors. 
    \item Second, there is an feature learning procedure using skip-gram model.
    \item Third, we use a neighbor-based feature propagation by propagating the node’s embedding to its neighbors(no sampling here)  with a parameter r.
\end{enumerate}
\begin{algorithm}[tb]
\caption{No-walk sampling strategy}
\label{alg:No-walk}
\textbf{Input}: G(V,E) \\
\textbf{Parameter}: Limit number of sampling neighbors num \\
\textbf{Output}: neighbors
\begin{algorithmic}[1] 
\STATE Initialize neighbors to Empty
\FOR{$v \in V$}
\STATE N(v) = neighbor(v)
\STATE Shuffle(N(v))
\WHILE{len(N(v)) less than num}
\FOR{$n \in N(v)$}
\STATE N(v).extend(neighbor(n))
\ENDFOR
\ENDWHILE
\STATE sentence = [v] + N(v)[:num]
\STATE neighbors.append(sentence)
\ENDFOR
\STATE \textbf{return} neighbors
\end{algorithmic}
\end{algorithm}
\subsection{No-walk sampling strategy}
When sampling node’s neighbors, the one-hop neighbors have the  priority, if they don’t reach the limit, we sample from the two-hop neighbors until either the limit is reached or the the two-hop neighbors have been all involved. In practice by our implementation, the fixed number should be proper with the average degree of the network to make sure skip-gram model can learn a sufficient representation of the nodes and their neighbors.The complexity of our strategy is O(n), which makes our strategy very efficient and more trivially parallelizable compared to the existed methods. 
Another advantage our strategy have over the other methods is that our strategy pays more attention on neighborhood relationship representation. To illustrate this, assuming we got a input like $[c, n_1, n_2, n_3, n_4, n_5, n_6]$ where the c is the center node and the $n_1-n_4$ are its one-hop neighbors and $n_5,n_6$ are its two-hop neighbors, and we set window size equals 7. When given this input, the skip-gram model not only learn center node’s one-hop, two-hop neighborhood relationship representation, but also n1’s one-hop, two-hop, three-hop neighborhood relationship representation. And it go furthest to four-hop for $n_5 and n_6$,  because they may be four-hop neighbor to each other with all conditions taken into consideration. To summarize, the diversity and abundance of neighborhood relationship representation is guaranteed  and our sampling mechanism makes us away from learning those neighborhood relationship which is too far and likely to be meaningless. And this advantage differs us from DeepWalk, node2vec, word2vec etc, when given a long “sequence” and large window size.

\begin{algorithm}[tb]
\caption{Skip-gram model}
\label{alg:Skip-gram}
\textbf{Input}: neighbors
\begin{algorithmic}[1] 
\FOR{$neighbor \in neighbors$}
\FOR{$c \in neighbor$}
    \STATE $J(\Phi) = -log Pr(c|\Phi(neighbor))$
    \STATE $\Phi = \Phi - \alpha \frac{\partial J}{\partial \Phi}$
\ENDFOR
\ENDFOR
\end{algorithmic}
\end{algorithm}

\begin{algorithm}[tb]
\caption{Neighbor-based feature aggregation and propagation}
\label{alg:feature smooth}
\textbf{Input}:  G(V,E),
	Propagation rate r,
	Raw matrix of node representations Mat\textunderscore r,
	Aggregating method (average, attention etc...) \\
\textbf{Output}: smoothed matrix of node representations Mat\textunderscore s

\begin{algorithmic}[1] 
\FOR{$v \in V$}
\STATE raw\textunderscore emb = Mat\textunderscore r(v) \\
\STATE N(v) = neighbor(v) \\
\STATE agg\textunderscore emb= aggregate(embedding of Mat\textunderscore r(N(v))) \\
\STATE smo\textunderscore emb = raw\textunderscore emb * (1-r) + agg\textunderscore emb*r \\
\STATE Mat\textunderscore s.append(smoothed\textunderscore emb)
\ENDFOR
\end{algorithmic}
\end{algorithm}

\subsection{Position-ignoring Skip-gram model}
Skip-gram model has been originally developed in NLP and is still playing import roles on word embedding and graph embedding. 
We find the skip-gram very suitable for network representation learning. First, the order independence assumption captures a sense of “nearness” that is provided by BFS. And the order of neighbor sequence makes no difference to the objective which is exactly the network needs, because our sampling strategy generates unordered neighbors. From where we stand, the sentences in skim-gram based NLP models can be viewed as position-based neighbors list. Although developed to deal with euclidean structure data, skip-gram model goes beyond euclidean structure and works in non-euclidean structure data too. 
This characteristic makes it effective and efficient in graph embedding works.The basic Skip-gram formulation defines co-occurrence probability using soft-max function which is impractical. A computationally efficient approximation of the full soft-max is the negative sampling]. Negative sampling is defined by the object:

\begin{equation}
\resizebox{.99\linewidth}{!}{$
    J(\Theta) = log \sigma ({v_{w_O}} ^ {\prime\mathrm{T}}v_{w_I}) + \sum_{i=1}^k E_{{w_i} \sim P_n(w)}[log \sigma (-{v_{w_i}} ^ {\prime\mathrm{T}}v_{w_I})]
$}
\end{equation}%

\subsection{Neighbor-based feature aggregation and propagation}
There is a technical method to reuse the structure information to help to generate the  graph embedding. Here  we use a neighbor-based feature aggregation and propagation to propagate the node’s feature information to its (one-hop) neighbors as well as aggregate the feature information from its neighbors. Parameter {\em r} control how much information the node shares with neighbors. This method only has influence on the  one-hop neighbors, but it be expanded two-hop and more through multiple {\em iterations}.

Our implement aggregates features from neighbors,and propagates features to the same neighbors for undirected network. As directed network, we aggregate from indegree neighbors and propagate to outdegree neighbors. As for weighted network, the aggregation and propagation is calculated in a weighted way. We add the aggregated features to the raw feature by a propagation rate {\em r} which is set default 0.1 and can be fine-tuned.

This methods is inspired from self-attention mechanism proposed by Transformer~\cite{vaswani2017attention}. Self-attention mechanism maps one variable-length sequence of symbol representations ($x_1, ..., x_n$) to another sequence of equal length ($z_1, ..., z_n$), and for each $z_i$ in  ($z_1, ..., z_n$), $z_i$ is a weighted sum of ($x_1, ..., x_n$). The weights is calculated by a attention function. Here we simplify this to a average of the neighbors’ embedding, as another aggregation method choice. Although  self-attention mechanism is originally developed to deal with euclidean structure data like language sentences, it goes beyond euclidean structure and works in non-euclidean structure data as the skip-gram does. In NLP model, each input get a fixed number of words but in our neighbor-based feature aggregation and propagation. We  make it more flexible scale to large networks with millions of nodes in a few seconds. If we reconsider self-attention mechanism as a feature propagation method in a non-euclidean structure, we can see that self-attention mechanism consider each input sequence  ($x_1, ..., x_n$) as a full-connected network which means every node in the network has a link with the others. For each $x_i$ in the sequence, the other ($x_1,x_2,x_{i-1},x_{i+1},,,x_n$) are its neighbors and it aggregates and propagates information by self-attention mechanism. From this perspective, we can take the Transformer as a model for full-connected networks with a fixed number of nodes. Self-attention mechanism has been showing a promising work by Transformer and other models. We developed neighbor-based feature aggregation  propagation methods using self-attention mechanism, and the experiments following show its effectiveness.

\section{Experiments}
In this section we provide an overview of the chosen datasets and methods which we will use in our experiments.
\subsection{Experimental Datasets}
We conduct our experiments to evaluate the performance gain from two proposed tricks on four node classification datasets of Open Graph Benchmark (ogb)~\cite{DBLP:journals/corr/abs-2005-00687}, including ogbn-arxiv, ogbn-proteins,  ogbn-products, ogbl-ppa, ogbl-collab and ogbl-citation2. The statistics of each dataset is shown in Table1. 
Moreover, ogbn-arxiv, ogbn-products, ogbl-ppa, ogbl-collab and ogbl-citation2 all contain node features, and their evaluation metrics are listed below; ogbn-proteins only contains edge features, and its evaluation metric is ROC-AUC; ogbl-ppa, ogbl-collab and ogbl-citation2 are for Link property prediction task and their metrics are hits@100, hits@50 and Mean Reciprocal Rank (MRR) respectively. 

All these data have been split  into train, valid and test data by ogb. The dataset ogbn-arxiv is split by year. We propose to train on papers published until 2017, validate on those published in 2018, and test on those published since 2019. For ogbn-products, ogb use the sales ranking (popularity) to split nodes into training, validation and test sets. For ogbn-protein, ogb split the protein nodes into training, validation and test sets according to the species which the proteins come from. For ogbl-ppa, ogb provide a biological throughput split of the edges into training/validation/test edges. For ogbl-collab, it is split by year. For ogbl-citation2, ogb split the edges according to time, in order to simulate a realistic application in citation recommendation since 2019.

The reason we choose ogb-dataset is that most of the frequently-used networks before are extremely small compared to real-application networks with nodes  range from millions to billions. For example, the widely-used node classification datasets, CORA, CITESEER, and PUBMED~\cite{yang2016revisiting}., 2016] , only have 2,700 to 20,000 nodes; the popular graph classification datasets from the TU collection~\cite{yanardag2015deep} only contain 200 to 5,000 graphs, and the commonly-used knowledge graph completion datasets, FB15K and WN18~\cite{bordes2013translating}, only have 15,000 to 40,000 entities. 

The small datasets also make it hard to rigorously evaluate data-hungry models, such as Graph Neural Networks. In fact, the performance of GNNs on these datasets is often unstable and nearly statistically identical to each other, due to the small number of samples the models are trained and evaluated on.

\begin{table}
\centering
\begin{tabular}{lcccr}
\hline
Datasets  & Nodes & Edges & Classes & Metric \\
\hline
arxiv & 169,343	& 1,116,243	& 40 & acc \\
proteins & 132,534	& 39,561,252 & 2 & auc \\
products & 2,449,029 & 61,859,252 & 47 &  acc \\
ppa & 576,289 & 30,326,273 & link & hits@100 \\
collab & 235,868 & 1,285,465 & link & hits@50 \\
citation2 & 2,927,963 & 30,561,187 & link & mrr \\

\hline
\end{tabular}
\caption{The experiment ogb data information}
\label{tab:ogb}
\end{table}

\subsection{Baseline Methods}
Our experiments evaluate the feature representations obtained through neighbor2vec on standard supervised learning tasks: label classification for nodes and link prediction for edges. For following tasks, we evaluate the performance of neighbor2vec against the following feature learning algorithm.
\begin{enumerate}
    \item DeepWalk: This approach learns d-dimensional feature representations by simulating uniform random walks. The sampling strategy in DeepWalk can be seen as a special case of node2vec with p = 1 and q = 1.

    \item node2vec: This approach is similar to DeepWalk. The difference comes from the sampling strategy which DeepWalk use a uniform random walk but  node2vec set up two parameter p and q to control the priority of DSF or BSF  at 2nd order random walk.

\end{enumerate}

\subsection{Experiments Setup}
To facilitate the comparison between our method and the relevant baselines, we use the exact same datasets and we split the data in the same way as ogb  does. We train model in the given train data and test in the given test data. We repeat this process 10 times, and report the average performance in terms of the mapping metric. 

For all baselines we use a 3 layers Muti-Layer Perception MLP for classification. We present results for baselines from related papers or official results, if it is available. Following ogb rules, we evaluate neighbor2vec with 10 runs, without fixing random seed.
In addition, we also present two classical node level neighborhood encoding algorithms, GCN~\cite{kipf2016semi} and GraphSAGE~\cite{hamilton2017inductive}, to demonstrate the effectiveness of the proposed framework. For GCN and GraphSAGE, we just copy the results from ogb official leader board.

\subsection{Label Classification}
Overall, the proposed neighbor2vec consistently outperforms all baselines in terms of given metrics at all three datasets. When we look at dataset individually (Table 2),  the performance of two baselines get  percent scores of accuracy over 70 but below 71 and neighbor2vec performances 71.79, a lightly up to the others on ogbn-arxiv. Moreover, the proposed neighbor2vec outperforms both GCN and GraphSAGE with a tiny advantage, which scores 71.49 and 71.74 respectively.

With the scale of dataset grows, neighbor2vec shows a more outstanding advantage over the others. On dataset  ogbn-proteins which has a lot of links, our proposed neighbor2vec shows a huge improvement up to 79.62, 11 percent higher than the DeepWalk and node2vec. Also, the proposed neighbor2vec  outperforms both GCN and GraphSAGE with a big advantage, which scores 72.51 and 77.68 respectively. On dataset ogbn-products which has 2.45 million nodes and 61.86 million edges, the proposed neighbor2vec is the only that scores more than 80 percent, and it outperforms both the baseline methods and the two classical GNNs. We also see that progation method has contributed exciting lifts over five datasets.

\begin{table}
\centering
\begin{tabular}{lccc}
\hline
Model & Arxiv & Proteins & Products \\
\hline
DeepWalk & 70.01 & 68.12 & 70.37 \\
node2vec & 70.07 & 68.81 & 72.49 \\
GCN & 71.49 & 72.51 & 75.64 \\
GraphSAGE & 71.74 & 77.68 & 78.29 \\
Neighbor2vec(no progation) & 70.53 & 77.47 & 78.06 \\
Neighbor2vec & \textbf{71.79} & \textbf{79.62} & \textbf{80.36} \\
\hline
gain & 1.72 & 10.81 & 7.87 \\
\hline
\end{tabular}
\caption{Label classification results}
\label{tab:label cls}
\end{table}

\begin{table}
\centering
\begin{tabular}{lccc}
\hline
Model & PPA & Collab & Citation2 \\
\hline
DeepWalk & 23.03 & 50.37 & 83.24 \\
node2vec & 22.26 & 48.88 & 61.41 \\
GCN & 18.67 & 44.10 & 84.74 \\
GraphSAGE & 16.55 & 48.10 & 82.60 \\
Neighbor2vec(no progation) & \textbf{31.63} & 49.58 & 84.28 \\
Neighbor2vec & \textbf{31.63} & \textbf{51.14} & \textbf{85.84} \\
\hline
gain & 8.60 & 0.67 & 2.60 \\
\hline

\end{tabular}
\caption{Link prediction results}
\label{tab:link pre}
\end{table}

\begin{figure*}[tb]
\centering
\includegraphics[width=18cm, height=5.0cm]{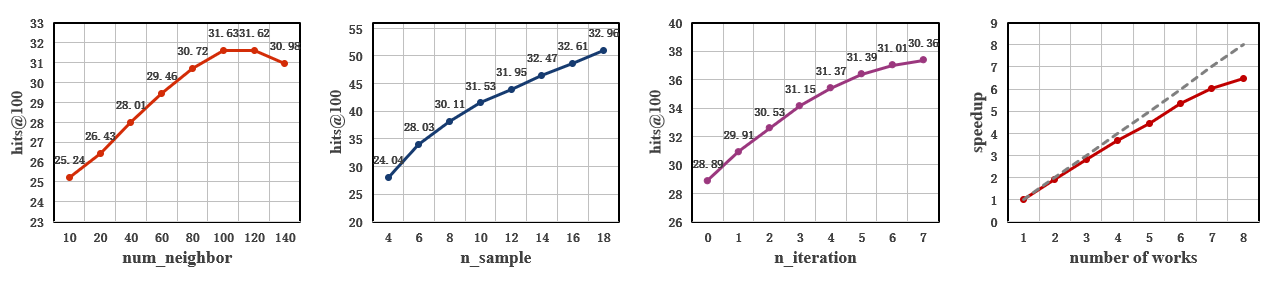}
\caption{Scalability of neighbor2vec}
\label{fig: Parameter sensitivity}
\end{figure*}

\subsection{Link Prediction}
In link prediction, we are given a network with a certain fraction of edges removed, and we would like to predict these missing edges. To facilitate the comparison between our method and the relevant baselines and claasical GNNs, we use three datasets ogbl-ppa, ogbl-collab and pgbl-citation2, and we split and evaluate the data as ogb  does. 

Just  like in label classification experiment, we still use a 3 layers muti-layer perception MLP for link prediction. Following ogb rules, we evaluate neighbor2vec with 10 runs, without fixing random seed. In addition, we also present two classical node level neighborhood encoding algorithms, GCN and GraphSAGE, to demonstrate the effectiveness of the proposed framework. We present results for baselines and GNNs from related papers or official results, if it is available.

We summarize our results of link prediction in table 3, from which we can easily draw a general conclusion that the learned feature representation for node pairs significantly outperform the baseline methods and the classical GNNs on all three datasets. On the dataset ogbl-ppa, the proposed neighbor2vec outperforms the DeepWalk and node2vec with huge gain up to 8.60 percent and 9.37 percent respectively, and it achieves  a hits@100 score to 31.63 percent, which is far over the two classical algorithms GCN and GraphSAGE. On ogbl-collab, neighbor2vec outperforms the DeepWalk and node2vec with light gain up to 0.67 percent and 2.26 percent respectively, and it achieves  a hits@50 score to 52.15 percent. Also  neighbor2vec  outperforms both GCN and GraphSAGE with a big advantage, which scores 44.10 and 48.10 respectively. On ogbl-citation2, neighbor2vec outperforms the DeepWalk and node2vec with gain up to 2.60 percent and 24.43 percent respectively, and it achieves  a MRR score to 85.84 percent. Also  neighbor2vec  outperforms both GCN and GraphSAGE with a small advantage, which scores 84.74 and 82.60 respectively. It seems both the GCN and GrapgSAGE do not work well on these link prediction task. And we see that the proposed neighbor2vec has a general as well as big advantage over the DeepWalk and node2vec over all tasks and dataets.

In summary, neighbor2vec learns a more stable and effective node embeddings than current state-of-the-art methods, as measured by multi-label classification performance as well as link prediction.  Moreover, sometimes it outperforms the classical GNNs such as GCN and GraphSAGE at different tasks and different networks. 

\subsection{Parameter sensitivity}

The neighbor2vec algorithms have a lot of parameters and we conduct a sensitivity analysis of neighbor2vec to these parameters. Figure 1 shows the different performance of the neighbor2vec with different choice of parameters on the ogbn-ppa dataset. When testing one parameter, we set the other parameters the default values.

We examine how the node’s neighborhood parameters (number of samples $n$-$sample$, number of neighbors $n$-$neighbor$) affect the performance. We observe that performance tends to saturate once the dimensions of the representations reaches around 100. Similarly, increasing both the number of walks and neighbors will improve the performance.

We also test the affect of the number of iterations of feature propagation $n$-$iteration$. The conclusion we draw from the figure is that the iteration of feature propagation improves the performance from 0 - 5, and the performance goes down with the iteration increases. This is reasonable because the growing iteration brings neighbor’s information in the beginning and brings in  the over-smoothing problem after several iterations. The propagation rate is a little bit like the learning rate in a supervised learning algorithm. These results demonstrate that we are able to learn meaningful latent representations for nodes after only a small number of iterations.

\subsection{Scalability}

We talk about scalability of our proposed algorithm here. To test for scalability, we learn node representations using neighbor2vec with default parameter values at dataset ogbn-products. Similiar to the DeepWalk and the node2vec, neighbor2vec can be parallelized by using the same mechanism.

We observe that neighbor2vec scales linearly with increase in number of nodes generating representations for 2.45 million nodes. We run experiments with the default parameters with different number of threads, i.e., 1, 2, 3, 4, 5, 6, 7, 8.  In general, we find that both methods achieve acceptable sublinear speedups as both lines are close to the optimal line. Overall, the proposed neighbor2vec is efficient and scalable for large-scale networks with millions of nodes.

\section{Discussion and Conclusion}
In this paper, we studied unsupervised feature learning in networks as a neighborhood relationship representation  optimization problem. And we studied to find an effective and  efficient way to process unsupervised feature learning at non-euclidean structure data. We treated skip-gram model with negative sampling as a classical non-euclidean data-matching algorithm and introduce a novel neighbor-based feature aggregation and propagation mechanism for unsupervised feature learning inspired  by the self-attention mechanism which is a feature propagation method at  a full-connected networks. These perspectives give us multiple advantages.

The first advantage of the proposed algorithm lies in their proper consideration and accommodation of the network’s structure and  property. We drop the random or guided walks which is developed by DeepWalk and node2vec respectively. Instead, we propose the BFS-based sampling strategy which focus only on the near neighborhood representations. From where we stand, we think that the walk-based algorithms do not sufficiently enough represent the neighborhood relationship for the network. Because they sample only a little near neighbors in a walk. Also they are very easily to learn the nose representation if given a large window size (which is set default 5). 

The second advantage of the proposed algorithm is that we introduced a novel neighbor-based feature aggregation and propagation mechanism. In our implements, we aggregate the features from the neighbors of the central node and add to its original features. 

Combination of these two advantages gives a effective yet efficient way on how to learn a good represent of network, while having minimal computational complexity. Our experiments have demonstrated that the proposed algorithm outperforms some classical GNN frameworks such as GCN and GraphSAGE in all of the tasks and datasets. Our future work in the area will focus on making full use of these  advantages, using our methods to improve network unsupervised learning as well as supervised learning, and strengthening the theoretical justifications of the method.

\section*{Acknowledgments}
We are thankful to Siyue Liu, Huayin Pan, Wei Lu, Xiumei Yang, Shaoyan He, Hongyang Wang for their enlightening suggestions as well as the anonymous reviewers for their helpful comments.


\bibliography{ijcai22}

\end{document}